\documentclass[lettersize,journal]{IEEEtran}
\usepackage{amsmath,amsfonts}
\usepackage{algorithmic}
\usepackage{algorithm}
\usepackage{array}
\usepackage[caption=false,font=normalsize,labelfont=sf,textfont=sf]{subfig}
\usepackage{textcomp}
\usepackage{stfloats}
\usepackage{url}
\usepackage{verbatim}
\usepackage{graphicx}
\usepackage{cite}
\usepackage{xcolor}
\usepackage{textgreek}
\usepackage[normalem]{ulem}

\begin{document}



\title{70 MW-level picosecond mid-infrared radiation generation by  difference frequency generation in~AgGaS$_2$, BaGa$_4$Se$_7$, LiGaSe$_2$, and LiGaS$_2$   }


\author{Michal Jelínek,
Milan Frank,
Václav Kubeček,
Ondřej~Novák,
Jaroslav Huynh,
Martin Cimrman,
Michal~Chyla,
Martin Smrž,
Tomáš Mocek

\thanks{Manuscript received October 10, 2024}  
\thanks{Corresponding author: Michal Jelinek: michal.jelinek@fjfi.cvut.cz }
\thanks{Michal Jelínek, Milan Frank, and Václav Kubeček are with Czech Technical University in Prague, Faculty of Nuclear Sciences and Physical Engineering, Brehova 7, Prague, Czech Republic}
\thanks{Ondřej~Novák, Jaroslav Huynh, Martin Cimrman, Michal~Chyla, Martin Smrž, and Tomáš Mocek are with HiLASE Centre, FZU - Institute of Physics of the Czech Academy of Sciences, Za~Radnicí 828, Dolní Brezany, Czech Republic}

\thanks{This work was co-funded by the European Union and the state budget of the Czech Republic under the project LasApp CZ.02.01.01/00/22\_008/0004573
and by "Center of Advanced Applied Sciences" (No. CZ.02.1.01/0.0/0.0/16\_019/0000778). }
}


\markboth{IEEE PHOTONICS JOURNAL,~Vol.~XX, No.~XX, October~2024}%
{Shell \MakeLowercase{\textit{et al.}}: A Sample Article Using IEEEtran.cls for IEEE Journals}


\maketitle

\begin{abstract}
Comparative study of nonlinear crystals for picosecond difference frequency generation in mid-IR is presented. 
Nonlinear crystals of AgGaS$_2$, BaGa$_4$Se$_7$, LiGaSe$_2$, and LiGaS$_2$ were studied.
In order to investigate the dependence of efficiency on the crystal length, three sets of crystals with lengths of  2, 4, or 8 mm were tested.
The developed tunable DFG system was driven by the 1.03~\textmu m, 1.8~ps, Yb:YAG thin-disk laser system operated at the~repetition rate of 10 or 100~Hz. 
As the best result, picosecond mid-IR pulses at a~wavelength of $\sim$7~\textmu m with the energy up to 130~\textmu J corresponding to the peak power of $\sim$72~MW were generated using the 8~mm long LiGaS$_2$ crystal. 
Using the BaGa$_4$Se$_7$ crystal, DFG tunability in the~wavelength range from 6~up~to~13~\textmu m was achieved.

\end{abstract}


\section{Introduction}

Sources of coherent radiation in a mid-infrared (IR) spectral range ($\sim$3-20~\textmu m) are interesting for many possible \mbox{applications} in vibrational spectroscopy, medicine (soft tissue ablation), biology (interaction with tissue or analysis), high-harmonic generation, attosecond science, atmospheric studies, etc.~\cite{VodopyanovWiley20, York08}.
Laser radiation around $\sim$3~\textmu m can be generated by well-developed Er$^{3+}$-doped lasers.  
Laser-diode pumped, quasi-continuous wave (QCW) lasers with $>$50~W of average output power are commercially available. Several works are also devoted to a~development of Er$^{3+}$:ZBLAN fiber lasers in continuous wave (CW) or Q-switched regime~\cite{Goya19,Sojka20}.
Nevertheless, not all temporal regimes are readily available and mainly stable mode-locked operation for ultrashort pulse generation is still challenging.

Going farther into the infrared region, laser radiation can be generated directly by some special solid-state lasers based on Fe$^{2+}$ or Dy$^{3+}$ active ions.
Lasers based on Dy$^{3+}$:PbGa$_2$S$_4$ can generate radiation at  narrow spectral lines around 4.3 or 5.4~\textmu m~\cite{Jelinkova16}.
Fe$^{2+}$-based lasers can provide generally continuously-tunable radiation in a~wide spectral range \mbox{because} of strong interaction of the electronic states with lattice vibrations.
Pumping of such lasers is generally not trivial because the Fe$^{2+}$ absorption band
is located  around 3~\textmu m. 
Moreover,  upper-state level lifetime at room temperature is in the order of hundreds of nanoseconds indicating special requirements  on pumping sources.
Therefore, Er$^{3+}$-based lasers, usually operated in a~Q-switched regime, are usually used as pumping sources of Fe$^{2+}$-based lasers.
Host materials such as ZnSe or ZnS were studied extensively and the~tuning range of $\sim$3.7 to 5.3~\textmu m was presented~\cite{MARTYSHKIN21,Fjodorow21}.
It has to be noted that the Fe$^{2+}$ lifetime as well as fluorescence spectra and resulting tunability range are  strongly influenced by the active material temperature.
The spectral tuning range can be extended using other Fe$^{2+}$ host materials such as CdTe or Cd\textsubscript{1-x}Mn\textsubscript{x}Te and pumping of such lasers by other Fe$^{2+}$:ZnSe lasers~\cite{DOROSHENKOFeCdMnTe18}.
For example, smooth tunability of a Fe$^{2+}$:CdTe laser in the~range $\sim$4.5 to 6.8~\textmu m was demonstrated~\cite{Frolov20}.
Concerning other temporal regimes, CW  Fe$^{2+}$:ZnSe laser  generation was achieved and mode-locked regime in the sub-picosecond time domain was also successfully reported~\cite{Pushkin20}. 
Some of these lasers became commercially available recently.

Moreover, there are commercially available and highly developed gas lasers based on CO or CO$_2$ generating within the wavelength ranges of 5.2–6.0~\textmu m or 9.1-10.9~\textmu m, respectively. These lasers can be operated in various temporal regimes from CW to ultrashort pulse generation~\cite{Endo06, IoninCO}. 
Free-electron lasers represent another type of very versatile and  broadly-tunable mid-IR radiation sources, but they suffer from complexity and costs~\cite{BrauFEL}. 
These devices generate in a~unique temporal regime. 
The  generated macropulse with a~duration of several \textmu s   consists of a~$\sim$GHz train of micropulses with $\sim$1~ps duration.
Semiconductor (quantum cascade) lasers present other interesting and applicable alternatives with CW output power at a~level of~several Watts~\cite{Pecharroman17, Slivken19}. \mbox{Generation} of low average power sub-picosecond pulses was also presented recently~\cite{Taschler21}. There are also other alternative approaches for generation in mid-IR, for example, hydrogen Raman lasers~\cite{Antognini05}.

An extensive group of applicable and highly versatile sources is based on  nonlinear optical parametric processes.
This concept has  a great  advantage in  possibility of output wavelength tunability in a~wide range given by the phase-matching conditions.
There are generally several \mbox{approaches}: optical parametric generation (OPG), optical parametric \mbox{amplification} (OPA), optical parametric oscillation (OPO), and difference frequency generation (DFG) that can be considered as a special type of OPA. These individual methods are commonly combined in some multi-stage systems (OPG+OPA, OPG+DFG etc.) enabling the generation of longer wavelengths or higher energies.  While OPG, OPA, and DFG can be essentially considered as single pass, OPO systems are based on multi-pass round-trip inside an optical resonator. Therefore, quality of an output beam generated by OPOs is generally better, usually close to a~fundamental mode, and also the operation threshold is lower. In contrast to OPG or OPA systems, peak powers and energies generated by OPOs are orders of magnitude smaller.


DFG process, also known as parametric down-conversion,  allows  conversion of high-energy, visible or near-IR laser radiation into a~mid-IR region or even further to THz frequency band~\cite{VodopyanovWiley20, PetrovPQE15,Leitenstorfer,Makhlouf}. 
Concerning more closely on mid-IR region, this concept has been proven in a~wide temporal range from CW~\cite{Insero16} to pulsed regime with pulse duration from nanoseconds down to femtoseconds~\cite{Su22}. 
As~DFG is a~three-wave mixing nonlinear process, the nonlinear crystals have to be non-centrosymmetric, i.e. present the second order nonlinearity $\chi^{(2)}$. A~choice of the nonlinear crystal depends on several factors and the desired wavelength belongs to the most important. Oxide crystals, as for example BBO (BaB$_2$O$_4$), KTP (KTiOPO$_4$), or LiNbO$_3$ are widely used.  Moreover, many of these crystals can be periodically-poled in order to increase conversion efficiency. On the other hand, their transparency range is limited, i.e. up to $\sim$2.9~\textmu m in the case of BBO,  up to $\sim$4.3~\textmu m for KTP, or up to $\sim$5.5~\textmu m for LiNbO$_3$~\cite{Pires15,Smrz17,refSNLO}.

In order to develop efficient frequency converters into wavelengths longer than 5~\textmu m, non-oxide crystals should be employed because of their wider transparency range up to $\sim$20~\textmu m in some cases. Generally, sulphide-, selenide-, or phosphide-based crystals can be used and huge effort has been made to fabricate and investigate such crystals having acceptably high effective nonlinear optical coefficient ($>$5~pm/V) and reasonably high optical damage threshold~\cite{PetrovPQE15}. Quasi-phase-matched (QPM), orientation-patterned (OP) semiconductors as OP-GaAs present other interesting possibilities; nevertheless, dimensions of such structures are limited by the fabrication process.

Moreover, because of two-photon absorption at $\sim$1~\textmu m, materials such as GaAs or GaSe have to be usually driven at $>$1.5~\textmu m presenting a~limiting factor for available pumping sources~\cite{XuOpGaAsOL17, LiuGaSe}. A similar situation in terms of available pumping sources  applies  for crystalline ZnGeP$_2$ (ZGP) also.


Concerning more closely on DFG systems excited by pulses at the~wavelength of $\sim$1~\textmu m, many works were devoted to nanosecond pumping.
For example, 10-ns pulses with the peak power of 3~kW at 7.9~\textmu m were generated using a~BaGa$_4$Se$_7$ (BGSe) crystal~\cite{Liu21}.
In order to achieve orders of~magnitude higher peak power, excitation by shorter pulses in a~picosecond range can be used.

Several other works concerning this approach based on the BGSe crystals were published. Using this crystal driven by a~30~ps Nd:YAG laser, output energy of 140-230~\textmu J at 8-14~\textmu m with the~peak power of up to $\sim$7~MW has been achieved~\cite{YangBGSe20}.  Moreover, the highest peak power of $\sim$27~MW at 3.9~\textmu m (energy of 830~\textmu J in 30~ps pulses) was generated under similar pumping conditions~\cite{YangBGSe13}.
Using LiInS$_2$ (LIS) crystal driven by a~25~ps Nd:YAG laser, output energy of 350~\textmu J at 5.4~\textmu m
corresponding to the~peak power of up to $\sim$14~MW has been demonstrated~\cite{York08}.
Our group has also presented preliminary results on BGSe under 1.8~ps pumping in~\cite{JelinekBGSeOSA22}. Using LiGaS$_2$ and LiGaSe$_2$ crystals it was possible to generate up to 50~\textmu J energy level in the~4.6-10.8~\textmu m wavelength range with the~peak power of up to $\sim$2.5~MW under pumping by a~1.064~\textmu m, 20~ps Nd:YAG laser~\cite{Smetanin20}.

There is also a~growing number of works devoted to intra-pulse difference frequency generation (iDFG) in a~timescale of tens of femtoseconds~\cite{Tian21,Nakagawa23,Chen18,Bournet22,Carnio23}.
 This method is generally  versatile, simple, and robust.
 On the other hand, generated mid-IR pulse energy is comparatively low.
 In order to generate such pulses at $\sim$\textmu J energy level, it is useful to incorporate  amplifier stages.
  For example, a~developed system extended by two OPA stages based on LiGaS$_2$ enabling to generate 6~\textmu J, $\sim$100~fs pulses at $\sim$6~\textmu m was demonstrated~\cite{Nakagawa23}.

We have developed a system generating between these two temporal ranges, i.e. pulses in a picosecond scale.
Mid-IR pulses with the energy up to  130~\textmu J corresponding to the peak power of $\sim$72~MW were generated. 
Using various nonlinear crystals, tunability in the~wavelength range from 5 up to 13~\textmu m was achieved. 
Comparative study of  AgGaS$_2$ (AGS), BaGa$_4$Se$_7$ (BGSe), LiGaS$_2$ (LGS), or LiGaSe$_2$ (LGSe) with various lengths was performed. 
The developed DFG system was driven by the 1.03~\textmu m, 1.8~ps, Yb:YAG thin-disk based Perla~B laser system at the HiLASE centre \cite{Smrz17}. 

\section{Experimental setup}
\begin{figure*}[ht!]
\centering\includegraphics[width=13cm]{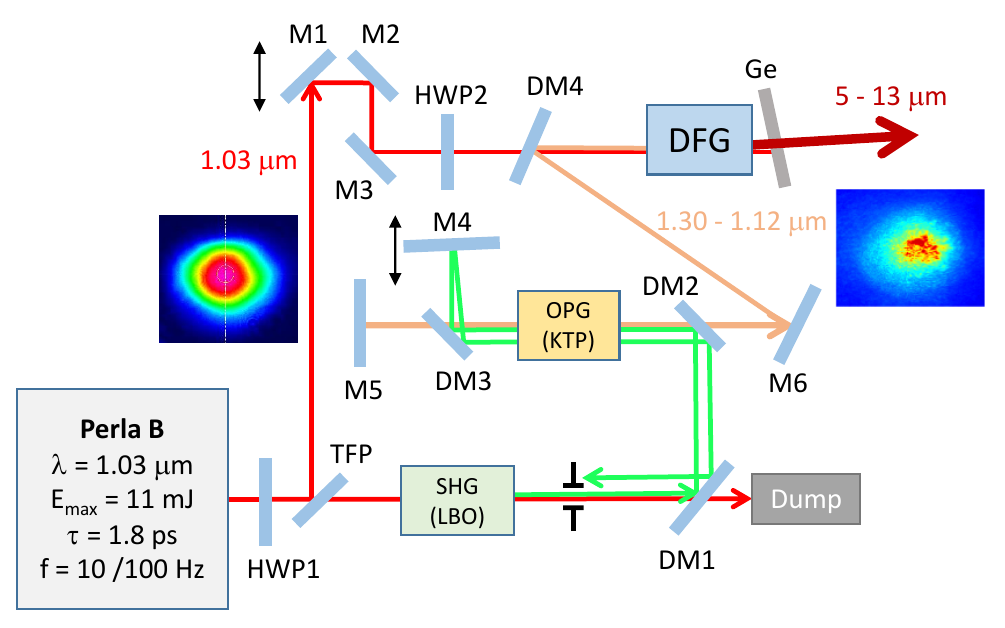}
\caption{Optical layout of the DFG system in collinear geometry. SHG: second harmonic generator, OPG: optical parametric generator, DFG: difference frequency generator. TFP: Thin film polarizer; HWP: half-wave plate @ 1030~nm; M1, M2, M3: HR @ 1030~nm; M4: HR @ 515~nm; M5 and M6: HR @ 1100-1300~nm. 
DM: dichroic mirrors:
DM1: HR @ 515~nm, HT @ 1030~nm; DM2 and DM3: HR @ 515~nm, HT @ 1100-1300~nm; DM4: HR @ 1120-1270~nm, HT @ 1030~nm. Ge: Germanium plate used to block pump as well as signal radiation. Insets: spatial profiles of pump and signal beams Additional lenses used for optimization of the beam spot sizes are not shown for better clarity.}
 \label{figSchematic}
\end{figure*}

The fundamental optical layout of the designed DFG system is presented in Fig.~\ref{figSchematic}. The principal idea of this setup was proposed in \cite{Yang15}.
Additional lenses were used in the setup in order to optimize beam spot size on the SHG, OPG, and DFG elements.

Notation in the order of pump – signal – idler ($\lambda$\textsubscript{pump}  $<$ $\lambda$\textsubscript{signal} $<$ $\lambda$\textsubscript{idler}) will be used in the following descriptions of three-wave mixing processes.
The 1.03-\textmu m laser with maximum pulse energy of 11~mJ was used as a~pump source for a double-stage frequency-converter into the mid-IR region.
According to the autocorrelation measurement, the pulse duration was 1.8~ps assuming sech$^2$ pulse shape.
The spectral bandwidth was $\sim$1.5~nm indicating that the pulse was not fully transform-limited.

The laser output energy was adjustable using a~motorized attenuator and the  pulse repetition rate was reduced from 1~kHz down to 10~Hz by a~pulse picker during most of the experiments. Furthermore, the setup with the 8~mm long BGSe crystal was tested at the highest output energy at the pump repetition rate of 100~Hz without any damage nor DFG output characteristics decrease. This result confirms that further mean output power scaling is possible.

The first stage was a~generator of a~signal wave for DFG based on a~double pass optical parametric generator - amplifier (OPG - OPA) system driven by a~second harmonic generation (SHG) stage.
Generated idler wave (tunable in a range of $\sim$1.12~to~1.30~\textmu m) was used as a~signal wave for the DFG stage. DFG was studied in several nonlinear crystals described in section~\ref{SectNlDFG}. The horizontally polarized pump beam (1.03~\textmu m) and vertically polarized signal beam ($\sim$1.12 to 1.30~\textmu m) were combined in the type II DFG crystal. A~delay stage formed by the mirrors M1-M2 was used in the pump beam for precise adjustment of the temporal overlap between the pump and signal beam. The HWP2 was used to adjust the correct polarization of the pump wave for DFG.

Details of the first stage of the frequency conversion system are following: The SHG was based on an LBO crystal with a~length of 5~mm, aperture of 5~$\times$~5~mm$^2$, cut at $\Theta$~=~90$^\circ$, \textphi~=~13$^\circ$, anti-reflection-coated (AR) at 515~nm and 1030~nm. 
During the DFG measurements described later, HWP1 was used to keep excitation pulse energy for the SHG crystal at a~constant level of 2.2~mJ. 
This value was limited by the damage threshold under given conditions.
Therefore, signal pulse energy for DFG was also kept at a~constant level while pump pulse energy for  DFG was varied during these measurements. 
Residual 1.03~\textmu m radiation was transmitted through DM1 into a~beam dump. SHG radiation at 515~nm served as excitation for the OPG-OPA system. It was based on a~KTP crystal with a~length of 5~mm, aperture of 10~$\times$~10~mm$^2$, cut at $\Theta$~=~73$^\circ$, \textphi~=~0$^\circ$, AR at 900–1400~nm, for o$\rightarrow$eo interaction (pump $\rightarrow$ signal, idler). During the first pass through the KTP, a~spectrally-broadband ($>$30~nm) and spatially-divergent conical idler beam was generated. It was transmitted through DM3 and reflected back into the KTP by M5. The pump beam at 515~nm was reflected by DM3 and further reflected back into the KTP by M4. M4 was placed on a~precise translation stage in order to adjust the temporal overlap between the pump and idler beams in the KTP during the second pass. The path of the back-reflected pump at 515~nm was slightly misaligned in the horizontal plane by M4 and therefore the back-reflected pump can be blocked by an iris placed in front of the SHG crystal. Using this second pass, the idler beam was amplified as well as spectrally and spatially narrowed. This idler beam was farther transmitted through DM2 and directed into the DFG crystal. The idler wave generated by this OPG-OPA system was tunable in the~wavelength range of ~1.12~to~1.30~\textmu m by slight rotation of the KTP crystal. The spectral full-width at half-maximum (FWHM) was ~10~nm and the maximum generated energy of $\sim$25~\textmu J was limited by the KTP damage threshold.
The spatial profile is presented in the inset of Fig.~\ref{figSchematic}. It has to be noted that there was also a~shorter (signal) wave generated around 0.93~\textmu m also but it was not used for further experiments and it was transmitted through DM4.

Nonlinear crystals used in the second stage of the frequency conversion are described in section~\ref{SectNlDFG} in detail.
The pump beam spatial profile was approximately Gaussian with an~average diameter of 2.8~mm (measured at 1/e$^2$) in the place of the DFG crystal. The calculated effective area of 3.1~mm$^2$ was used for further energy density calculations in accordance with the ISO standard 21254-1. A smaller pump beam may lead to higher conversion efficiencies at lower pump energies, but the main objective of this work was to use the maximum available pumping energy and generate as high mid-IR energy as possible without the DFG crystal damage. The signal beam diameter in the place of the DFG crystal was about 1.6~mm.

\section{ Nonlinear crystals for mid-IR  difference frequency generation } \label{SectNlDFG}
\begin{figure}[ht!]
\centering\includegraphics[width=9cm]{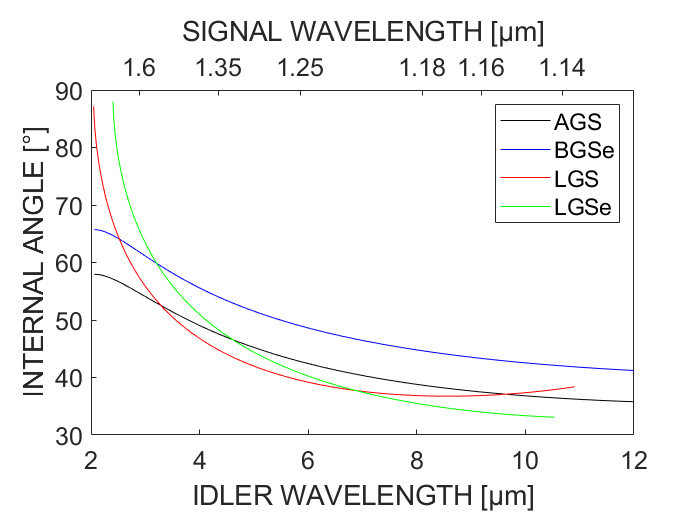}
\caption{ Calculated phase-matching curves of AGS, BGSe, LGS, and LGSe crystals for collinear setup in principal planes according to Tab.~\ref{tabCrystals}. }
\label{FigPM}
\end{figure}

DFG was investigated in  nonlinear crystals of AGS, BGSe, LGS, and LGSe.
Overview of all investigated samples is presented in Table~\ref{tabCrystals} together with fundamental crystallographic properties. Length of the samples was in the~range between 1 and 8~mm. Some of the crystals were anti-reflection (AR) coated at 1.2-1.5~\textmu m. Nonlinear characteristics of the crystals were calculated using SNLO~\cite{refSNLO}. The crystals were cut at the angles ($\Theta$, \textphi) enabling access to a reasonably high effective nonlinear optical coefficient d\textsubscript{eff} for the selected type of interaction at the following central wavelengths: pump 1.03~\textmu m, signal 1.20~\textmu m, idler 7.27~\textmu m. It has to be noted that magnitudes of nonlinear optical coefficients used in SNLO for BGSe are somewhat uncertain and several papers presenting novel data were published recently~\cite{Zhao21,Guo22}. On the other hand, values of  d\textsubscript{eff} for AGS, LGS, and LGSe have been validated with the same results in many publications. Phase-matching curves are presented in Fig.~\ref{FigPM}.

\begin{table*}[h!]
\caption{ Overview of the investigated DFG crystal samples. }
\label{tabCrystals}

\begin{tabular}{l | c | c c c | c c c |  c c c}
\hline
 ~ & AGS & \multicolumn{3}{c|}{BGSe}  & \multicolumn{3}{c|}{LGSe} & \multicolumn{3}{c}{LGS} \\ \hline
 Anisotropy & neg. uniax. & \multicolumn{3}{c|}{biaxial} & \multicolumn{3}{c|}{biaxial} & \multicolumn{3}{c}{biaxial} \\
  Class & 42m & \multicolumn{3}{c|}{m} & \multicolumn{3}{c|}{mm2} & \multicolumn{3}{c}{mm2} \\
   DFG interaction type & 1 & \multicolumn{3}{c|}{1} & \multicolumn{3}{c|}{2} & \multicolumn{3}{c}{2} \\
    Pump-Signal-Idler & eoo & \multicolumn{3}{c|}{oee} & \multicolumn{3}{c|}{eoe} & \multicolumn{3}{c}{eoe} \\
     Principal plane & -  & \multicolumn{3}{c|}{XZ} & \multicolumn{3}{c|}{XY} & \multicolumn{3}{c}{XY} \\
      $\Theta$; \textphi & 34$^\circ$; 45$^\circ$ & \multicolumn{3}{c|}{45$^\circ$; 0$^\circ$} & \multicolumn{3}{c|}{90$^\circ$; 34$^\circ$} & \multicolumn{3}{c}{90$^\circ$; 39$^\circ$} \\
    d\textsubscript{eff}, pm/V & 10.5 & \multicolumn{3}{c|}{27} & \multicolumn{3}{c|}{9.2} & \multicolumn{3}{c}{5.6} \\
   Length, mm & 2 & 1 & 4 & 8 & 2 & 4 & 8 & 2 & 4 & 8 \\
Aperture, mm$^2$ & 8$\times$5 & 4$\times$4 & 5$\times$5 & 5$\times$6 & 4$\times$4 & 4$\times$4 & 4$\times$4 & 5$\times$5 & 4$\times$4 & 4$\times$4 \\
AR coating & no & no & no & yes & yes & yes & yes & no & no & no \\
\hline

\end{tabular}
\end{table*}

\begin{figure}[ht!]
\centering\includegraphics[width=9cm]{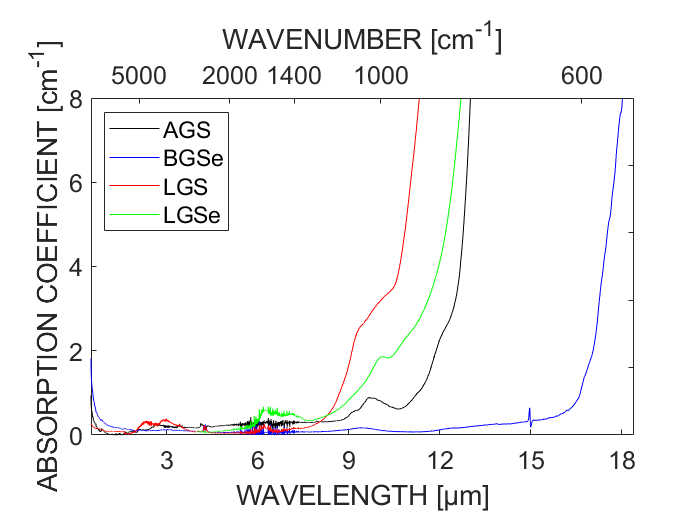}
\caption{ Spectral dependencies of net absorption coefficients for AGS, BGSe, LGS, and LGSe. }
\label{FigAbsorption}
\end{figure}

Unpolarized transmission spectra of these crystals were measured using a~spectrophotometer (Shimadzu UV-3600) and an FT-IR spectrometer (Nicolet). 
Fresnel reflections were subtracted, spectral dependencies of the net absorption coefficient  were calculated  and they are presented in Fig.~\ref{FigAbsorption}. 
Slight noise at $\sim$4.3~\textmu m is essentially caused by CO$_2$ absorption and noise around 6.4~\textmu m is caused by water vapor absorption. These data are in a~good agreement with SNLO~\cite{refSNLO}. It can be seen that the absorption coefficient increases significantly (higher than 2~cm$^{-1}$) above $\sim$9.0~\textmu m for~LGS, above $\sim$10.6~\textmu m for LGSe, above $\sim$11.9~\textmu m for AGS, and above $\sim$17~\textmu m for~BGSe.

\section{Characteristics of mid-IR  difference frequency generation  }
Laser-induced damage threshold (LIDT)  was measured \mbox{using}  same pump laser system but slightly different beam spot sizes in some cases.
Input surface damage was observed at pump pulse energy densities of approximately 0.16 and 0.21~J/cm$^2$ for the AR-coated BGSe and LGSe crystals, respectively.
In order not to damage other crystals, the energy density was kept slightly lower during the following DFG measurements.
Although these measured values are not fully comparable with the values published for the~uncoated crystals, they present a good estimation for further measurements.
The uncoated AGS and LGS crystals were not damaged by the pump energy density of 0.15 and 0.23~J/cm$^2$, respectively.

Mid-IR single pulse energy generated by the DFG system operated at~10~Hz was measured by a~sensitive energy probe (Coherent J10MB-LE).
When the system was operated at 100~Hz, the single pulse energy was measured by the energy probe and also mean power was measured by a~power meter (Thorlabs PM16-401). The single pulse energy was calculated from this power measurement and the values are in good agreement.

\begin{figure*}[ht!]
\centering\includegraphics[width=18cm]{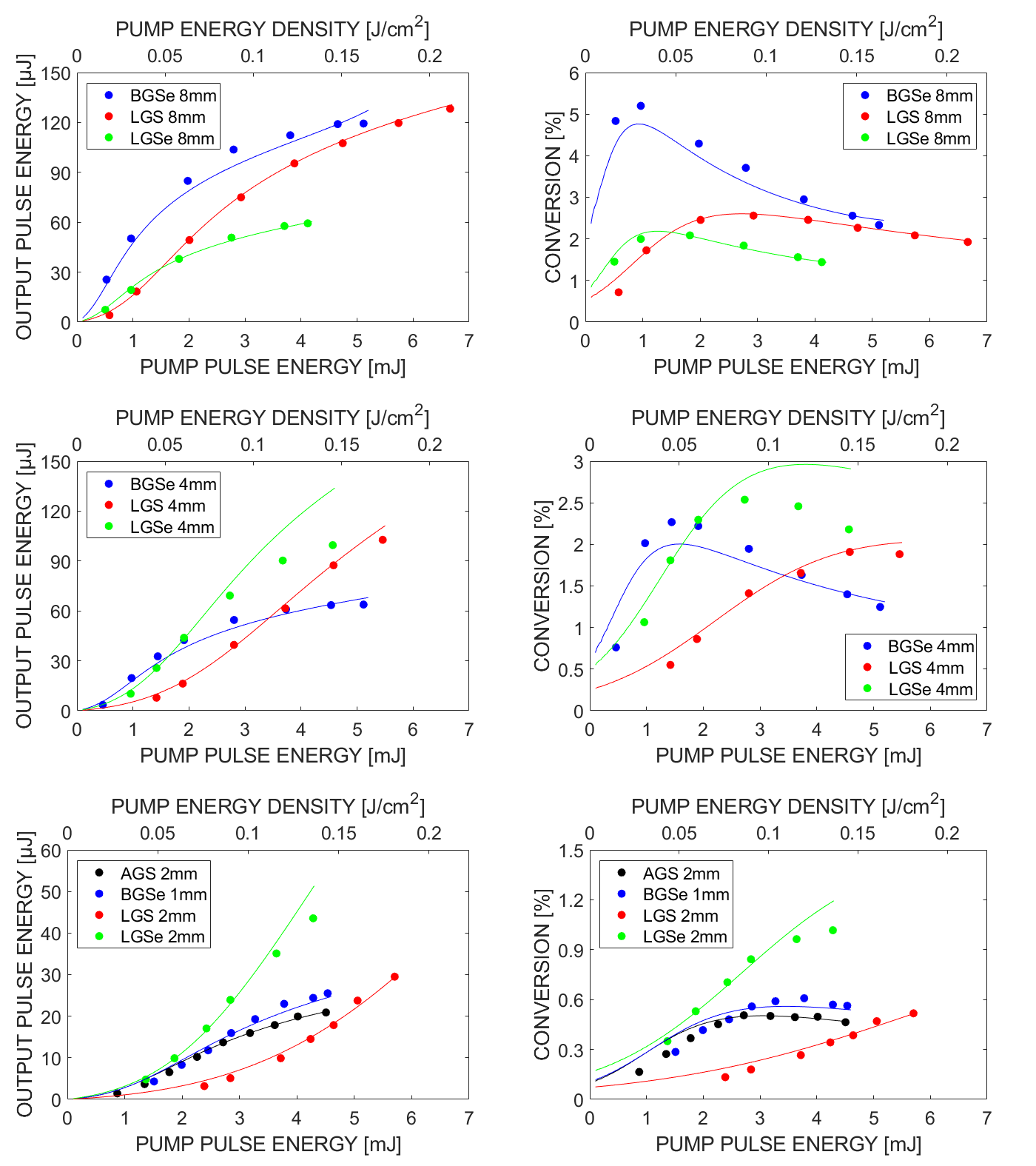}
\caption{ Mid-IR $\sim$7~\textmu m output energy (left) and conversion efficiency (right) achieved with 8~mm long DFG crystals (top row), 4~mm long DFG crystals (middle row), and 1 or 2~mm long DFG crystals (bottom row). Points: measured data; line: numerical modeling.}
\label{FigEnergyOverview}
\end{figure*}

Characteristics of generated pulse energy and energy conversion efficiency of the output radiation at the~wavelength of $\sim$7~\textmu m as a~function of the pump pulse energy for all investigated crystals are presented in Fig.~\ref{FigEnergyOverview}.
This wavelength was chosen because of the crystals cut as well as limited LGS transmission range.  
Germanium filter transmittance was taken into account, i.e. the presented values are in front of this filter. All crystals were tested up to the pump energy density close to the limit given by the LIDT value. It has to be also noted that the 8~mm long LGSe crystal was partially damaged during previous experiments that might negatively affect the output energy. The highest output energy of 130~\textmu J with the conversion efficiency of 2\% was achieved using the 8~mm long LGS crystal for the pump energy of 6.7~mJ corresponding to energy density of almost 0.2~J/cm$^2$. Assuming the output pulse duration comparable to the pump pulse duration of 1.8~ps, the peak power exceeded 72~MW level. 
It has to be noted that for  the highest pump level, the optimal signal energy  might be  higher, but it was limited by the OPG/OPA performance.  
Maximum conversion efficiency for this crystal was 2.5\% at~lower pump energy of 3~mJ. Using the 8~mm long BGSe crystal, the highest output energy was 120~\textmu J for pump energy of 5.1~mJ with the conversion efficiency of 2.3\%. It should be also noted that at low pump energy of 1~mJ, the output energy was about 50~\textmu J and the conversion efficiency reached 5\%. Using the 8~mm long LGSe crystal, the highest output energy was 60~\textmu J with the conversion efficiency of 1.4\%.
Using shorter crystals, the output energies were generally lower. In the case of 4~mm long crystals, the maximum output energy was $\sim$100~\textmu J for the LGS and LGSe crystals, and $\sim$60~\textmu J for the BGSe crystal. The conversion efficiencies were up to 2.5\% in the case of LGSe crystal for pump energy of 2.5~mJ. In the case of 1 or 2~mm long crystals, the maximum output energy was 45~\textmu J for the 2~mm long LGSe crystal. Using the 2~mm long LGS, 2~mm long AGS, and 1~mm long BGSe crystals, the maximum output energy was comparable at a~level of 25~\textmu J. The maximum conversion efficiency was 1\% for the 2~mm long LGSe crystal.
Shot-to-shot energy stability at high pump energy level was less than $\pm$5\%.

Presence of energy roll-off effect was observed for longer crystals or crystals with higher nonlinear coefficients. This phenomenon is probably given by re-conversion of the idler wave back into the pump and signal waves.
Solid lines in the graphs present results of numerical modeling based on SNLO~\cite{refSNLO} taking into account the spatial distribution with diffractive effects. It has to be noted that a shorter interaction length was used in this numerical modeling, taking into account non-optimal spatial or temporal overlap of the pump and signal beams. According to the simulation, in the ideal case for the 8~mm long crystals, the maximum output energy would be about 2~times higher.

The generated spectra in the mid-IR region were measured using a~single grating monochromator (Oriel model 77250 with 77302 grating covering the spectral range \mbox{4-16}~\textmu m, 75~lines/mm) together with a~cryogenically-cooled mercury cadmium telluride (MCT) photodetector (Judson-Teledyne J15D12, spectral range 3-13~\textmu m).  Spectral resolution was on the order of 30~nm. The signal wave at the~desired wavelength for DFG was tuned by the KTP crystal rotation, and the DFG crystal was rotated correspondingly in order to match the phase-matching condition.

\begin{figure*}[ht!]
\centering\includegraphics[width=18cm]{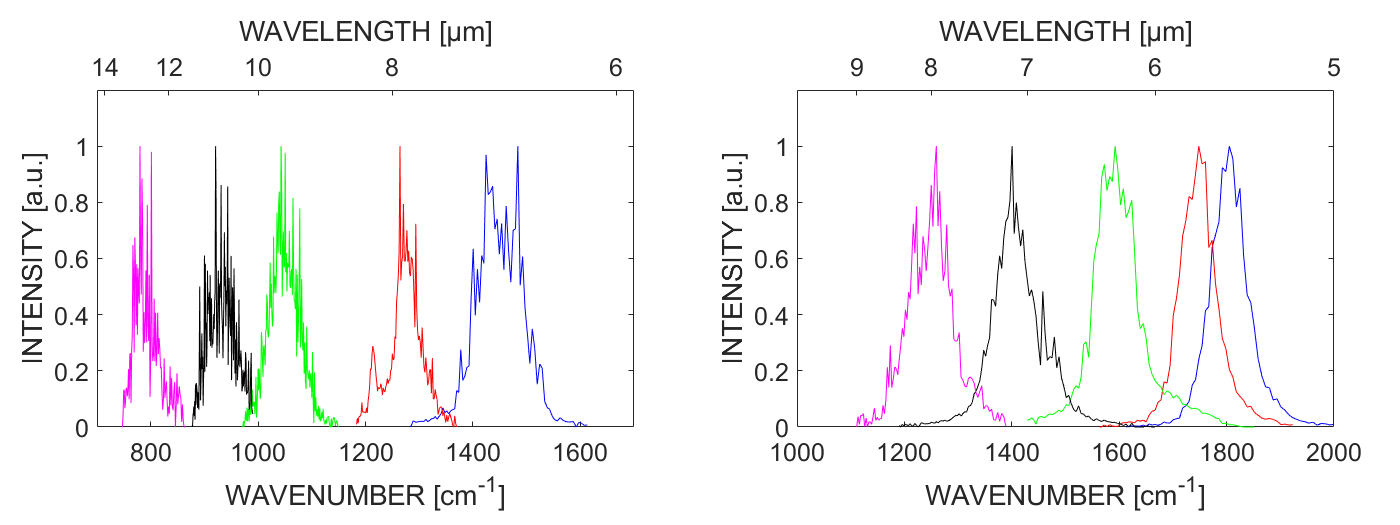}
\caption{ DFG spectral tunability using the BGSe crystal (left) and LGS crystal (right) at high pump energy.}
\label{FigTunability}
\end{figure*}

Examples of the spectra generated by DFG using the 8~mm long BGSe and LGS crystals are presented in Fig.~\ref{FigTunability}. In the case of the BGSe crystal, tunability in the~range from $\sim$6.5~to~13~\textmu m was achieved. Shorter wavelengths might possibly be generated, but effort has been made to investigate a~long-wavelength spectral part. The spectral width (FWHM) ranged from 80~cm$^{-1}$ at $\sim$7~\textmu m to 40~cm$^{-1}$ at $\sim$12.7~\textmu m. It has to be noted that it was not possible to measure longer wavelengths than $\sim$13~\textmu m because of the MCT photodetector sensitivity limit. Using the 8~mm LGS crystal, the overall tunability range was $\sim$5.3 to 8.3~\textmu m and the long-wavelength part was probably limited by the crystal transmittance. The spectral width was ranging from 70~cm$^{-1}$ at $\sim$5.5~\textmu m to 80~cm$^{-1}$ at $\sim$8~\textmu m. 

\section{Discussion and Conclusion  }
A comparative study of nonlinear crystals for picosecond difference frequency generation in the mid-IR was presented. Crystals of AgGaS$_2$, BaGa$_4$Se$_7$, LiGaSe$_2$, and LiGaS$_2$ with various lengths were used. The developed tunable DFG system was driven by the 1.03~\textmu m, 1.8~ps, Yb:YAG thin-disk laser system operated at a~repetition rate of 10 or 100~Hz. Picosecond mid-IR pulses at  a~wavelength of $\sim$7~\textmu m with energy up to~130~\textmu J corresponding to the peak power of $\sim$72~MW were generated using the 8~mm long LiGaS$_2$ crystal. Using various nonlinear crystals, DFG tunability in the~wavelength range from $\sim$6 up to 13~\textmu m was achieved.
In agreement with simulation results, higher conversion efficiency up to 5\% can be achieved for low pump energy densities using crystals with higher d\textsubscript{eff} (BGSe or LGSe).
On the other hand, at high pump energy densities approaching the crystal damage threshold, the conversion efficiency of 1 to 2\% is comparable for all crystals due to strong reconversion. Longer crystals (4 or 8~mm) are beneficial for higher achievable maximum generated energy for the investigated pump pulse duration.

Yb:YAG thin-disk pump laser allows further increase of the output pulse repetition rate up to 1~kHz  and therefore average output  power scalability of the whole DFG system.

 
%






\end{document}